\documentclass[conference]{IEEEtran}
\pdfoutput=1
\ifCLASSINFOpdf
\else
\fi

\usepackage{hyperref}       
\usepackage{url}            
\usepackage{booktabs}       
\usepackage{amsfonts}       
\usepackage{nicefrac}       
\usepackage{microtype}      
\usepackage{lipsum}
\usepackage{graphicx}
\usepackage{comment}
\usepackage{subcaption}
\usepackage{amsmath}
\usepackage{multirow}

\usepackage[utf8]{inputenc}
\begin{document}

\title{The Banking Transactions Dataset and its Comparative Analysis with Scale-free Networks*}

\author{
    \IEEEauthorblockN{Akrati Saxena\IEEEauthorrefmark{1}, Yulong Pei\IEEEauthorrefmark{1}, Jan Veldsink\IEEEauthorrefmark{2},Werner van Ipenburg\IEEEauthorrefmark{2}, George Fletcher\IEEEauthorrefmark{1}, Mykola Pechenizkiy\IEEEauthorrefmark{1}}
    \IEEEauthorblockA{\IEEEauthorrefmark{1}Eindhoven University of Technology, Eindhoven, The Netherlands 
    \\\{a.saxena,y.pei.1,g.h.l.fletcher,m.pechenizkiy\}@tue.nl}
    \IEEEauthorblockA{\IEEEauthorrefmark{2}Co\"{o}peratieve Rabobank U.A., Utrecht, The Netherlands
    \\\{jan.veldsink,werner.van.ipenburg\}@rabobank.nl}
}


%

\maketitle
\thispagestyle{plain}
\pagestyle{plain}

\begin{abstract}
We construct a network of 1.6 million nodes from banking transactions of users of Rabobank. We assign two weights on each edge, which are the aggregate transferred amount and the total number of transactions between the users from the year 2010 to 2020. We present a detailed analysis of the unweighted and both weighted networks by examining their degree, strength, and weight distributions, as well as the topological assortativity and weighted assortativity, clustering, and weighted clustering, together with correlations between these quantities. We further study the meso-scale properties of the networks and compare them to a randomized reference system. We also analyze the characteristics of nodes and edges using centrality measures to understand their roles in the money transaction system. This will be the first publicly shared dataset of intra-bank transactions, and this work highlights the unique characteristics of banking transaction networks with other scale-free networks.

\end{abstract}


%
\IEEEpeerreviewmaketitle

\section{Introduction}


The analysis of large-scale complex networks, including social networks, technological networks, biological networks, has offered in-depth insights to uncover and explore their evolution and understand several dynamic phenomena taking place on these networks \cite{barrat2008dynamical}. Some interesting examples, where complex networks have been used to provide solutions for various real-world problems, include understanding human communication pattern \cite{onnela2007analysis}, designing fair policies for overcoming pandemic \cite{atwood2019fair}, disrupting communities in terrorist networks \cite{miller2018discovering}, medicine development and treatment of complex diseases \cite{silverman2020molecular}, and so on. The analysis of bank transaction network will be beneficial for several research directions. 
\begin{itemize}
    \item It will help in understanding the flow of money at the microscopic level as well as how this contributes towards the macroscopic money transaction system.
    \item It will improve the downstream tasks in the financial domain with guidance from the topological perspective. Some representative tasks including fraud detection and user classification in transaction networks.
    \item It will shed light on financial simulator design. Existing simulators focus on individual customer behavior, while our analysis can provide complementary information about collective behaviors of group users.
\end{itemize}


The existing studies on financial network analysis focus on trade network or inter-bank transaction networks that show different patterns from intra-bank transaction data. Most of the studies on the financial domain, either make use of a financial simulator to generate synthetic transactions, e.g., PaySim~\cite{lopez2016paysim}, or concentrate on specific downstream tasks such as anomaly detection~\cite{pei2020subgraph}. In this paper, we aim to bridge the gap by presenting a detailed analysis of unweighted and weighted networks constructed from a data set of the money transactions of more than 1.6 million users of Rabobank.

This paper analyzes unweighted and weighted, large-scale, one-to-one bank transaction networks. We follow the `cookbook approach' for performing a systematic analysis of basic and more advanced network characteristics, which has also been followed in previous network analysis work \cite{onnela2007analysis}. We explain the characteristics of the dataset and networks in Section \ref{secdataset}. We study some of the basic network characteristics in Section \ref{secbasic}. We explore meso-scale structural properties of the network, including k-cliques and the coupling of link weight with the network topology in Section \ref{secmeso}. We briefly highlight centrality-based roles of nodes and edges in Section \ref{seccentrality}. We present a comparative analysis of the unique characteristics of the transaction network with other scale-free networks in Section \ref{seccompare} and, finally, discuss our findings and future directions in Section \ref{conclusion}.

Our main contributions are mentioned below.
\begin{enumerate}
    \item The anonymized banking transaction data of users. As per the best of our knowledge, this will be the first publicly available dataset of users' banking transactions. The dataset is available at \url{https://github.com/akratiiet/RaboBank_Dataset}.  
    \item We perform a detailed analysis of unweighted and weighted banking transaction networks and highlight the similarities and differences of these networks with other types of networks. As per the best of our knowledge, this is the first work to analyze intra-bank transaction network. We hope the shared data and this work will help further help in understanding the evolution of money flow in society. 
\end{enumerate}

\section{Dataset and Networks}\label{secdataset}
The dataset is collected from the Co\"{o}peratieve Rabobank U.A.\footnote{\url{https://www.rabobank.nl/}}, a Dutch multinational banking and financial services company. This dataset consists of bank accounts and transactions between them. For any pairs of accounts with one or more transactions, we also collected more information, including the numbers of transactions between two accounts and the total amount of money transferred from one account to another over a period of 11 years from 2010 to 2020. Thus, it can be organized into a transaction network, either unweighted or weighted. The data was shared for 1,624,030 bank accounts and 4127043 transactions based on (from\_account, to\_account) pair. 

We create the network from this data, having 1,624,030 nodes which are accounts, and 3,823,167 edges which represent that the respective users performed one or more transactions. Next, we identify the weakly connected components in the network, and it contains 723 connected components, where the largest weakly connected component contains 1,622,173 nodes and 3,821,514 edges. The information of connected components is: 1 component has 1,622,172 nodes, 3 components have 27, 15, and 13 nodes, and the rest of the components have less than ten nodes. The weakly largest connected component will be referred to as $G(V, E)$ where $V$ is the set of node and $E$ is the set of edges. In our analysis, we mainly consider the weakly largest connected component (WLCC) if not mentioned otherwise.

Considering the transaction information, we create two weighted networks (i) edge-weight is the total amount of money transferred between two accounts, and (ii) edge-weight is the total number of transactions between two accounts. The first network will be referred to as $G^T$, and the second will be referred to as $G^N$. If there are more than one rows for a (from\_account, to\_account) pair in the banking data, then the weights are added to compute the edge-weight in $G^T$ and $G^N$ networks.

Newman categorized the real-world scale-free networks into four main categories: (i) Biological, (ii) Information, (iii) Technological, and (iv) Social \cite{newman2003structure}. The banking transaction network falls under the category of \textit{Information Network}.

\section{Basic Network Characteristics}\label{secbasic}

\begin{figure*}[!]
    \centering
    \includegraphics[width=.9\linewidth]{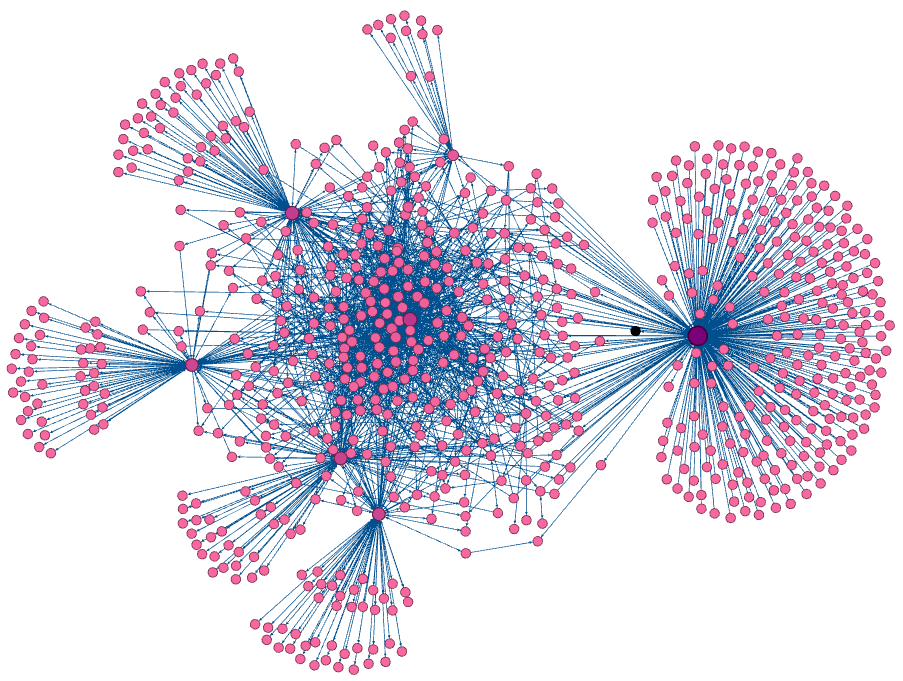}
    \caption{A sample subgraph extracted from the black node (chosen uniformly at random) till distance $l=3$.}
    \label{samplesubgraph}
\end{figure*}

We show a small subgraph of the network in Fig.~\ref{samplesubgraph} that is sampled using snowball sampling \cite{lee2006statistical}. The sample has been extracted from the network by picking a source node at random (shown with black color) and including all nodes in the sample that are within a topological distance of $l = 3$ from the source node. The node size and color are scaled based on the total degree (in-degree + out-degree) of the node. It appears from this figure that the network consists of a few hubs that are well connected with each other and small clusters of nodes as previously observed in other scale-free networks \cite{onnela2007analysis, newman2006structure}.

\begin{figure}
    \centering
    \includegraphics[width=\linewidth]{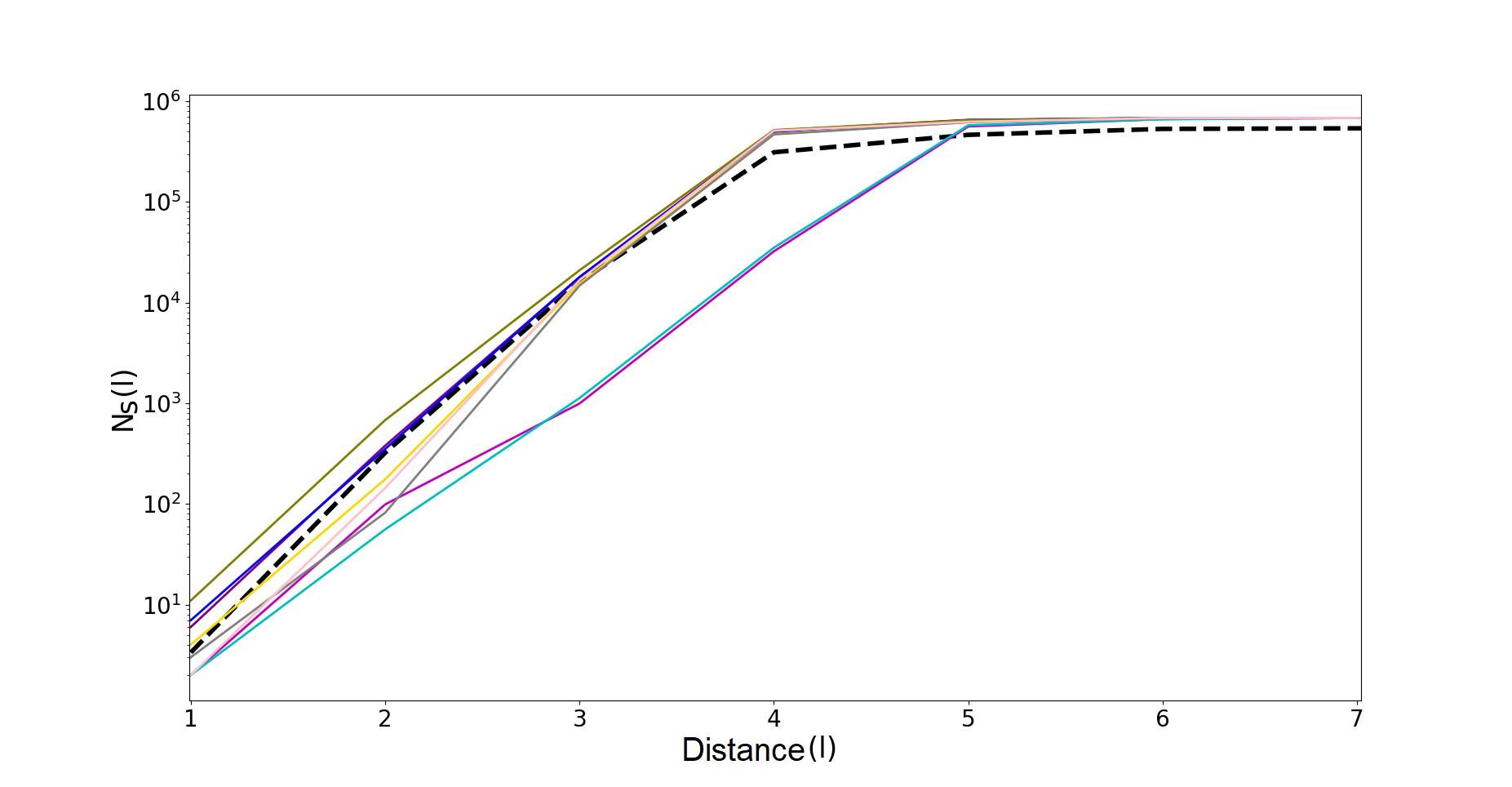}
    \caption{Average Number of nodes $(N_s(l))$ at distance $l$ as a function of distance $(l)$ for uniformly sampled $0.1\%$ source nodes (dashed black lines) and the number of nodes at distance $l$ for some random nodes (solid thin lines).}
    \label{fig:avg}
\end{figure}

To further understand network structure, we show in Fig.~\ref{fig:avg} the average number of nodes $N_s(l)$ as a function of distance $l$, where $N_s(l)$ represents the number of nodes within a distance $l$ from a given source node. The plots are shown for several choices of the source node using solid colored lines, and the average value of $N_s(l)$ for uniformly sampled $0.1\%$ source nodes is shown by the black dashed line. The $N_s(l)$ is computed for a small sample due to the high computational complexity. To a good approximation, we analyze that the curve follows the Boltzman equation $B+(A-B)/(1+(x/x_0)^p)$, where $A$, $B$, $x_0$, and $p$ are the parameters~\cite{cercignani1988boltzmann}.

\begin{figure}
    \centering
    \includegraphics[width=\linewidth]{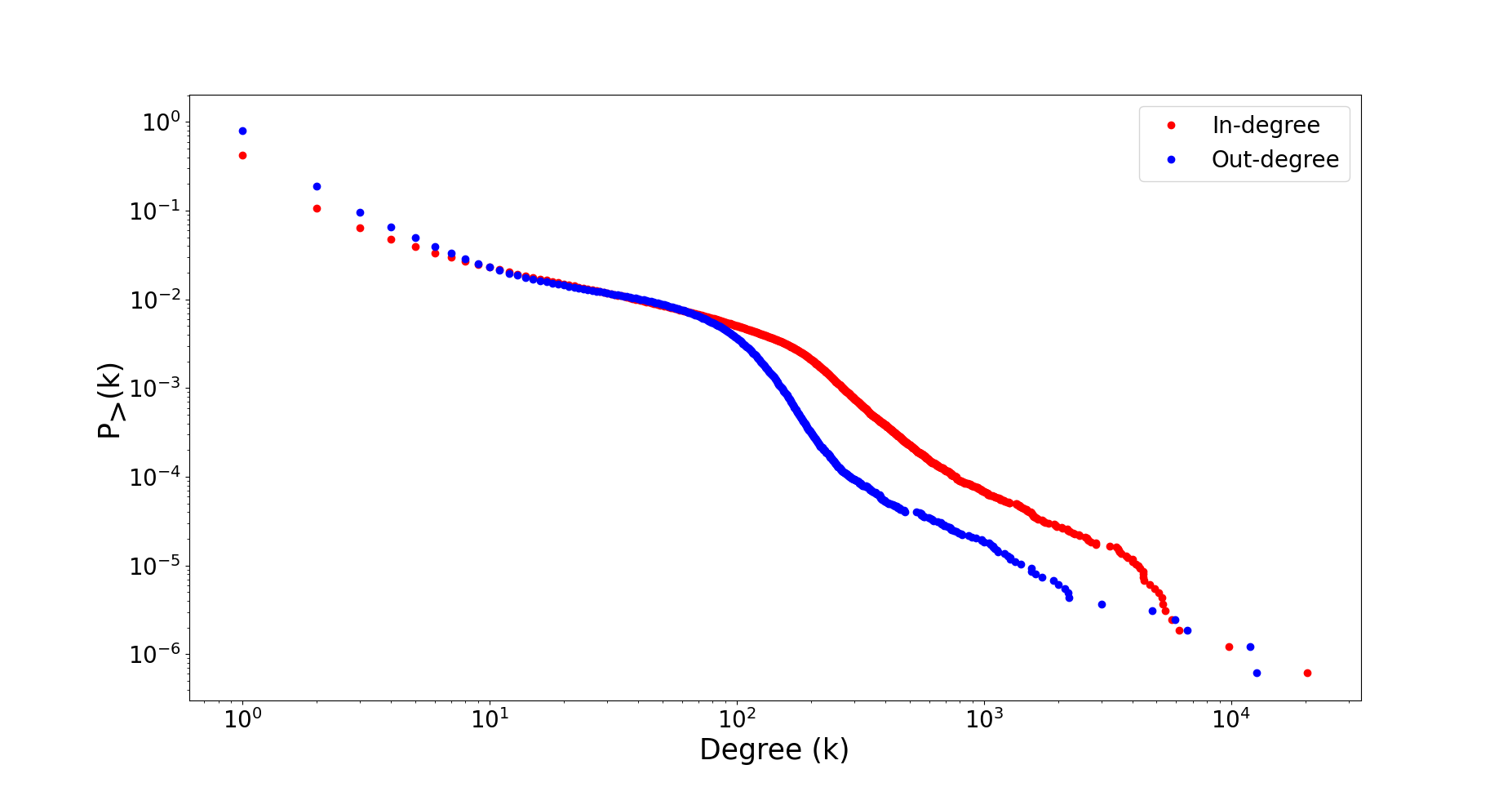}
    \caption{Cumulative In-degree and out-degree distribution.}
    \label{fig:cum_degree}
\end{figure}

We further analyze the macro-scale properties of the network. In Fig.~\ref{fig:cum_degree}, we show the cumulative in- and out-degree distribution of the weakly largest connected component (WLCC). The cumulative degree distribution $P_>(k)$ is defined as $P_>(k) =\int_{k}^{\infty}p(x) dx$, where $p(x)$ is the degree probability density function. The cumulative degree distribution $P_>(k)$ for the WLCC is approximated well by the power-law equation of the form $P(k) = c*(k)^{(-\gamma)}$, having the exponent $\gamma \approx 1.43$ and $1.13$ for the cumulative in-degree and out-degree distribution, respectively. Similar to the communication network~\cite{onnela2007analysis}, this network also has a rapidly decaying degree distribution where hubs are few, and therefore this network follows the properties of traditional scale-free networks such as anomalous diffusion.

\begin{figure}
    \centering
    \includegraphics[width=\linewidth]{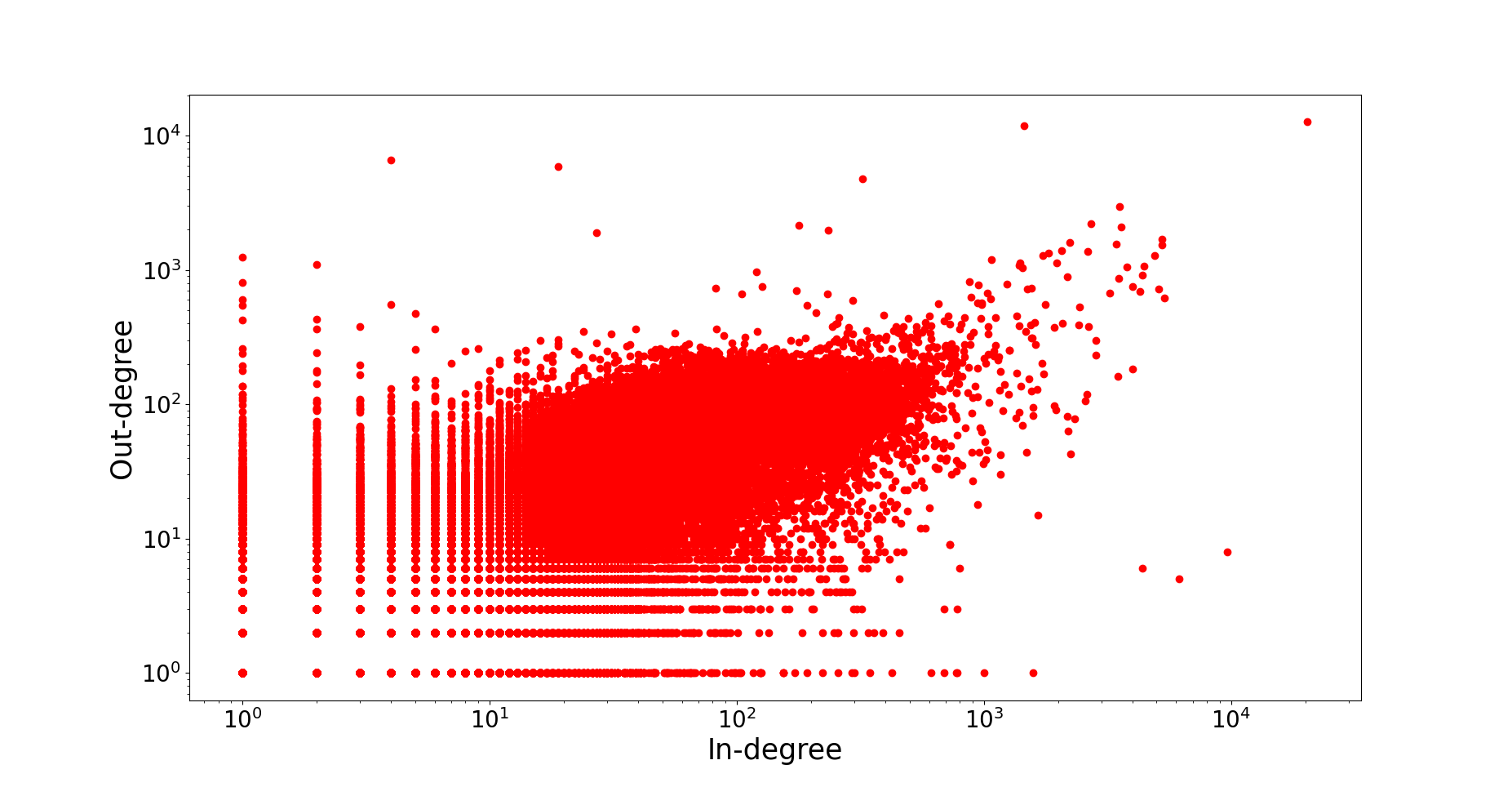}
    \caption{Out-degree versus In-degree for WLCC.}
    \label{figoutdegvsindeg}
\end{figure}

For a better understanding of the network, we plot the out-degree versus in-degree in Fig.~\ref{figoutdegvsindeg}; the plot shows that the out-degree is not correlated with in-degree and has the Spearman correlation coefficient $-0.15$. It indicates that there is no correlation between in-degree and out-degree. That is reasonable because most accounts that belong to regular customers have more transfer-out operations, e.g., purchasing products, than transfer-in operations, e.g., receiving salaries. 

\begin{figure}
    \includegraphics[width=\linewidth]{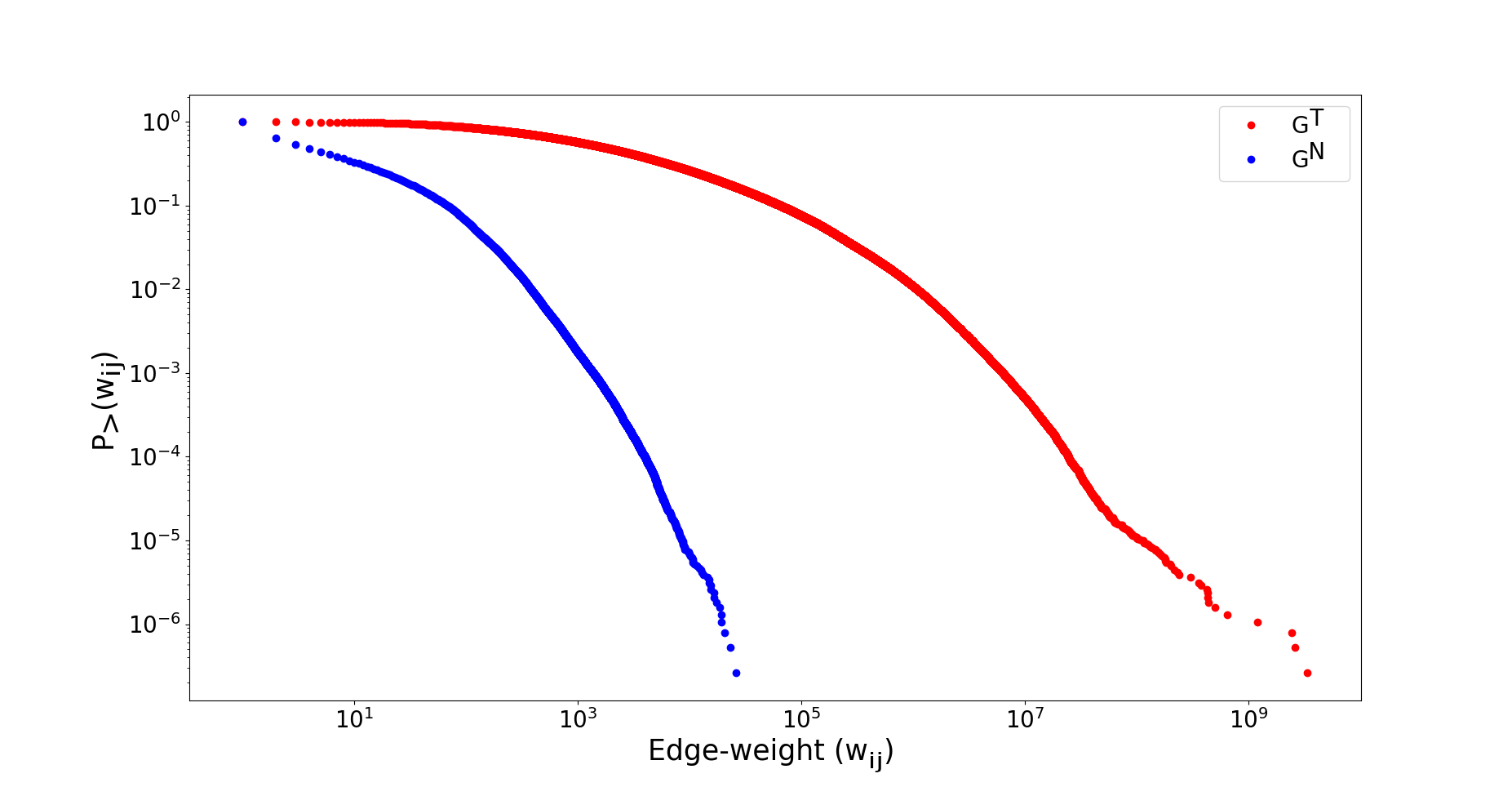}
  \caption{Cumulative edge-weight distribution for $G^T$ and $G^N$.}
  \label{edgeweightdist}
\end{figure}

\begin{figure}
    \centering
    \includegraphics[width=\linewidth]{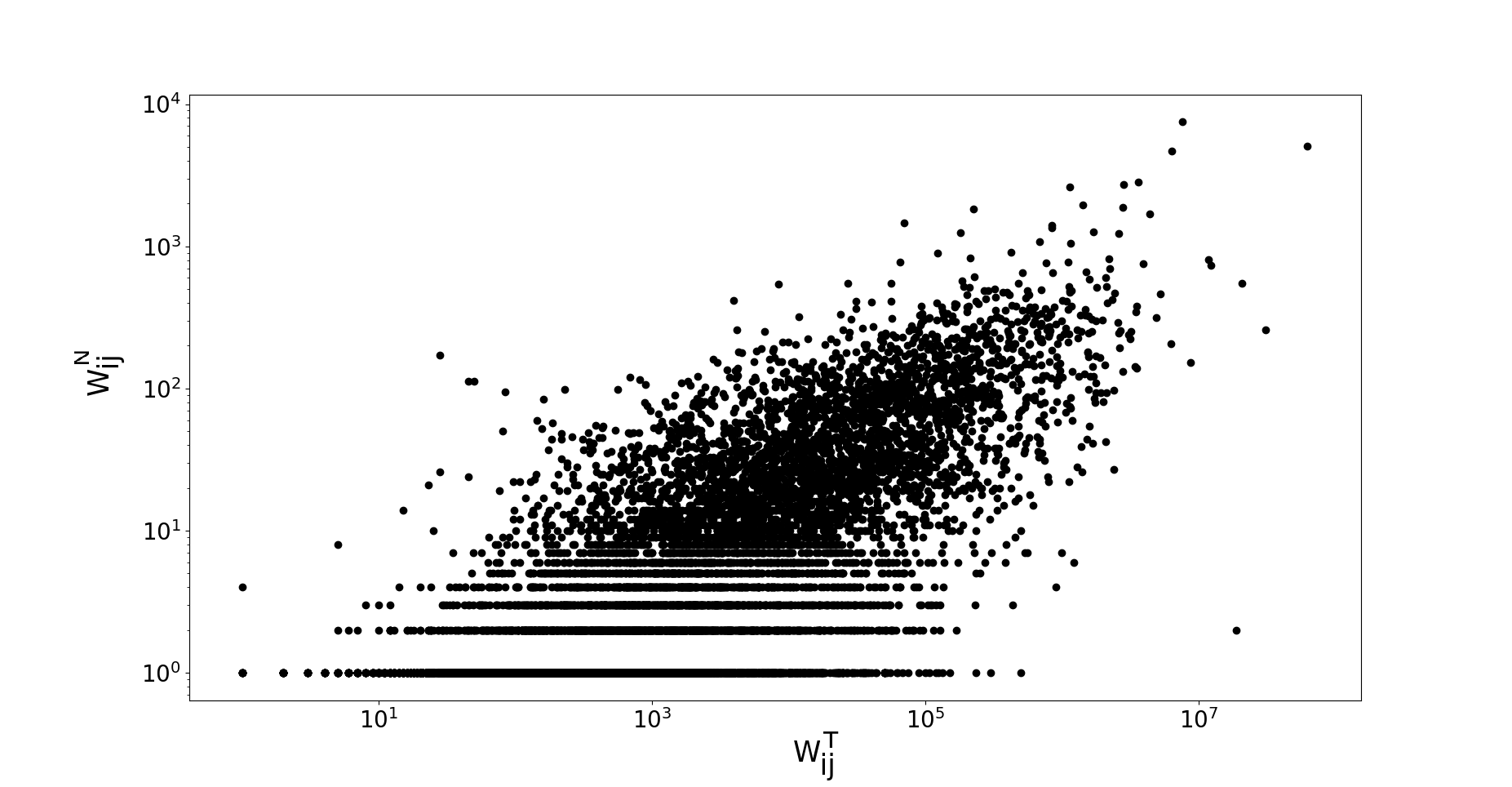}
    \caption{Correlation of edge-weights for randomly sampled 10000 edges in two networks $G^T$ and $G^N$. The two weights are clearly correlated having Spearman's coefficient 0.71.}
    \label{edge-weight correlation}
\end{figure}

By taking into consideration the transaction information, i.e., times of transactions and amount of money, we construct two weighted networks, $G^T$, and $G^N$. The cumulative edge-weight distributions are shown in Fig.~\ref{edgeweightdist}. We also analyze the correlation of edge-weights in both the networks that denote the dependency of the total transferred amount with the number of transactions. In specific, we show the correlation for randomly sampled 10000 edges in Fig.~\ref{edge-weight correlation}, and observe that the Spearman correlation coefficient of both types of edge-weights is 0.71. The results show that the users have a high correlation between the total amount transferred and the number of transactions for transferring that amount over time.

\begin{figure}
    \includegraphics[width=\linewidth]{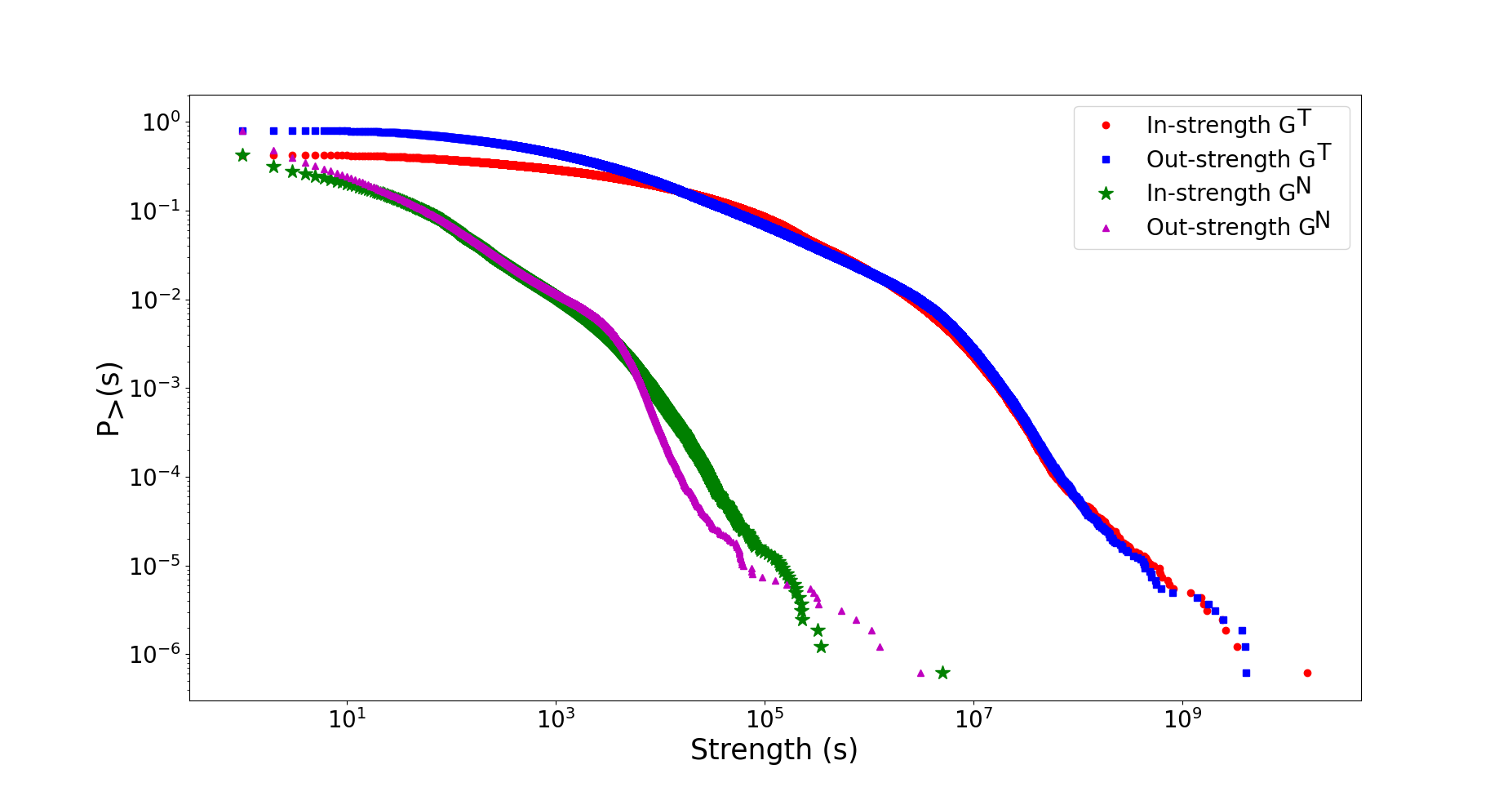}
  \caption{Cumulative In-strength and Out-strength distribution.} 
  \label{inoutstr}
\end{figure}

\begin{figure}
    \centering
    \includegraphics[width=\linewidth]{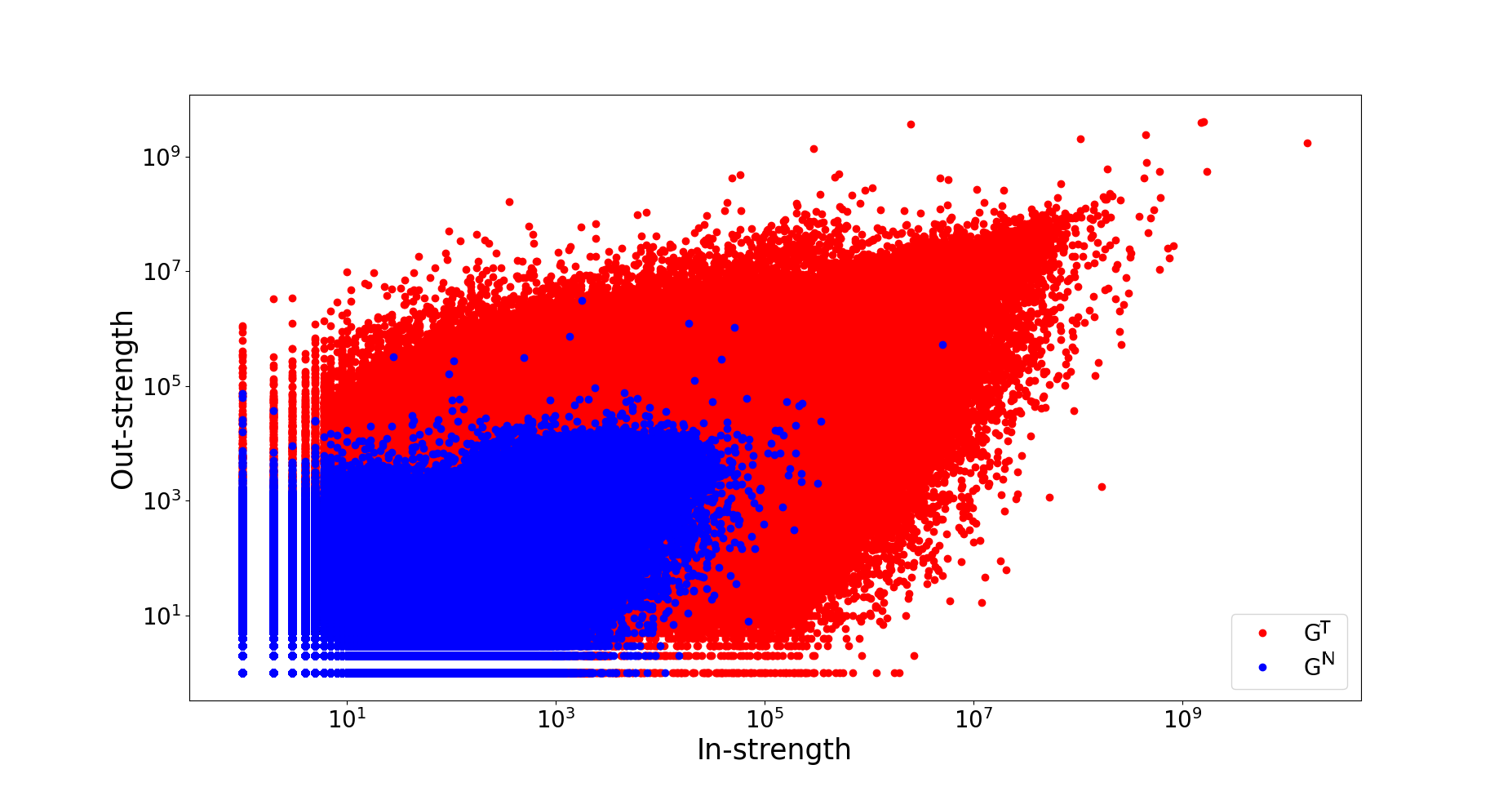}
    \caption{Out-strength versus In-strength for $G^T$ network in red color and for $G^N$ network in blue color.}
    \label{outstrvsinstr}
\end{figure}

The weights on edges contain richer information of transactions, so we compute the strength where the in-strength $s^{in}(i)$ and out-strength $s^{out}(i)$ of a node $i$ is the sum of weights of incoming edges and out-going edges, respectively. For example, the in-strength of a node $i$ is computed as, $s^{in}(i)=\sum_{(j,i) \in E}w_{ji}$, where $w_{ji}$ is the weight of an edge from node $j$ to node $i$. We investigate the cumulative in- and out-strength distributions of both weighted networks, and the plots are shown in Fig.~\ref{inoutstr}. As expected, both distributions follow the power law. The distributions of in-strength and out-strength are similar because in the process of data collection, we filter out special accounts such as gas stations that have much higher in-degrees than out-degrees. In Fig.~\ref{outstrvsinstr}, we show the out-strength versus in-strength for $G^T$ and $G^N$ networks having the Spearman correlation $-0.08$ and $-0.09$, respectively. We observe no correlation between out-strength and in-strength as expected, as most outgoing transactions are made to buy basic necessities for daily life, including grocery shopping or gas station payments, and these transactions have already been removed from the considered data. However, further detailed analysis will help in highlight the fraction of expenses for different types; this is out of scope for this work due to the unavailability of this data.

\begin{figure}[!]
  \begin{subfigure}{0.48\textwidth}
    \includegraphics[width=\linewidth]{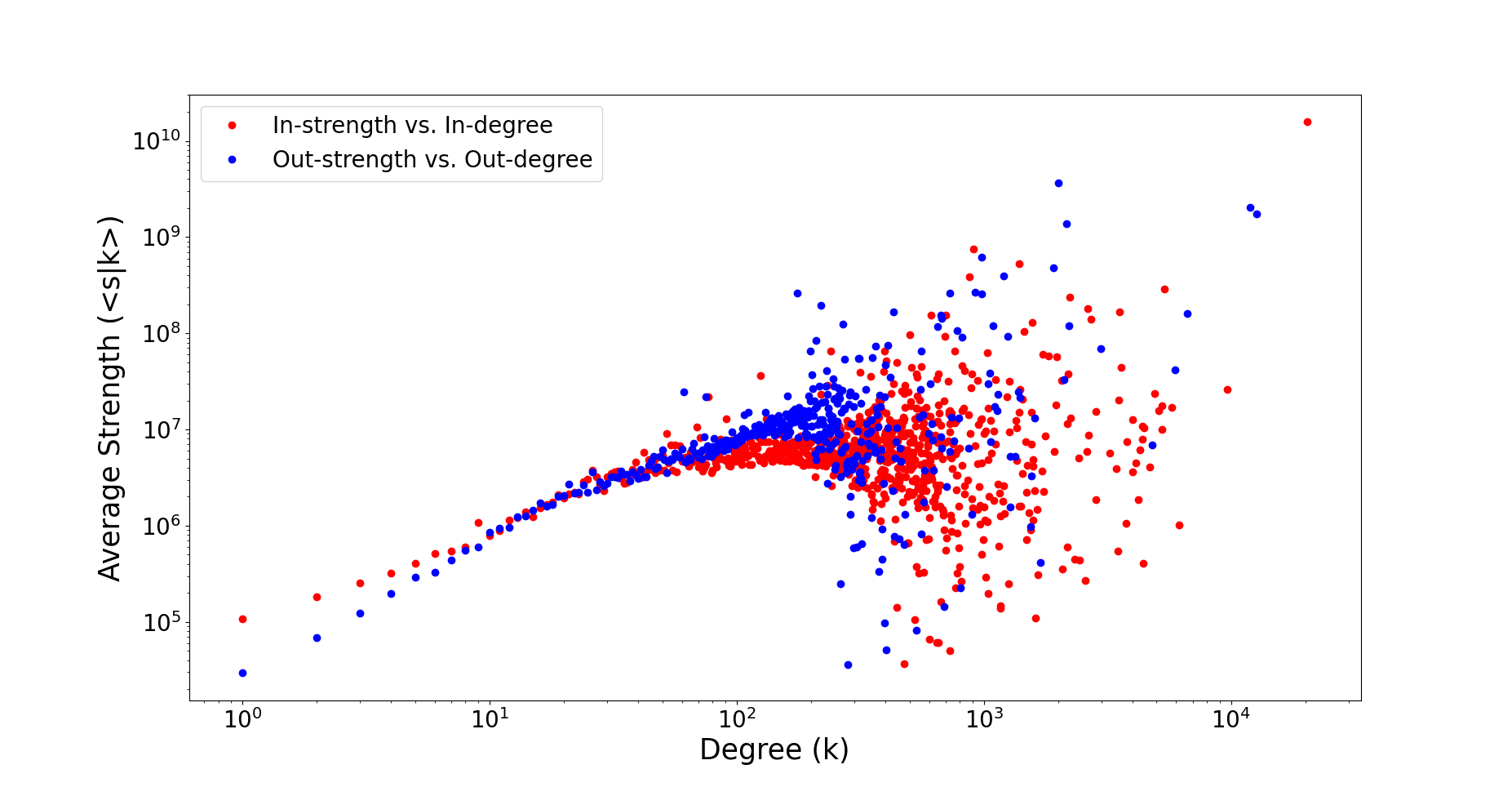}
    \caption*{(i) $G^T$}
  \end{subfigure}
  \hfill
  \begin{subfigure}{0.48\textwidth}
    \includegraphics[width=\linewidth]{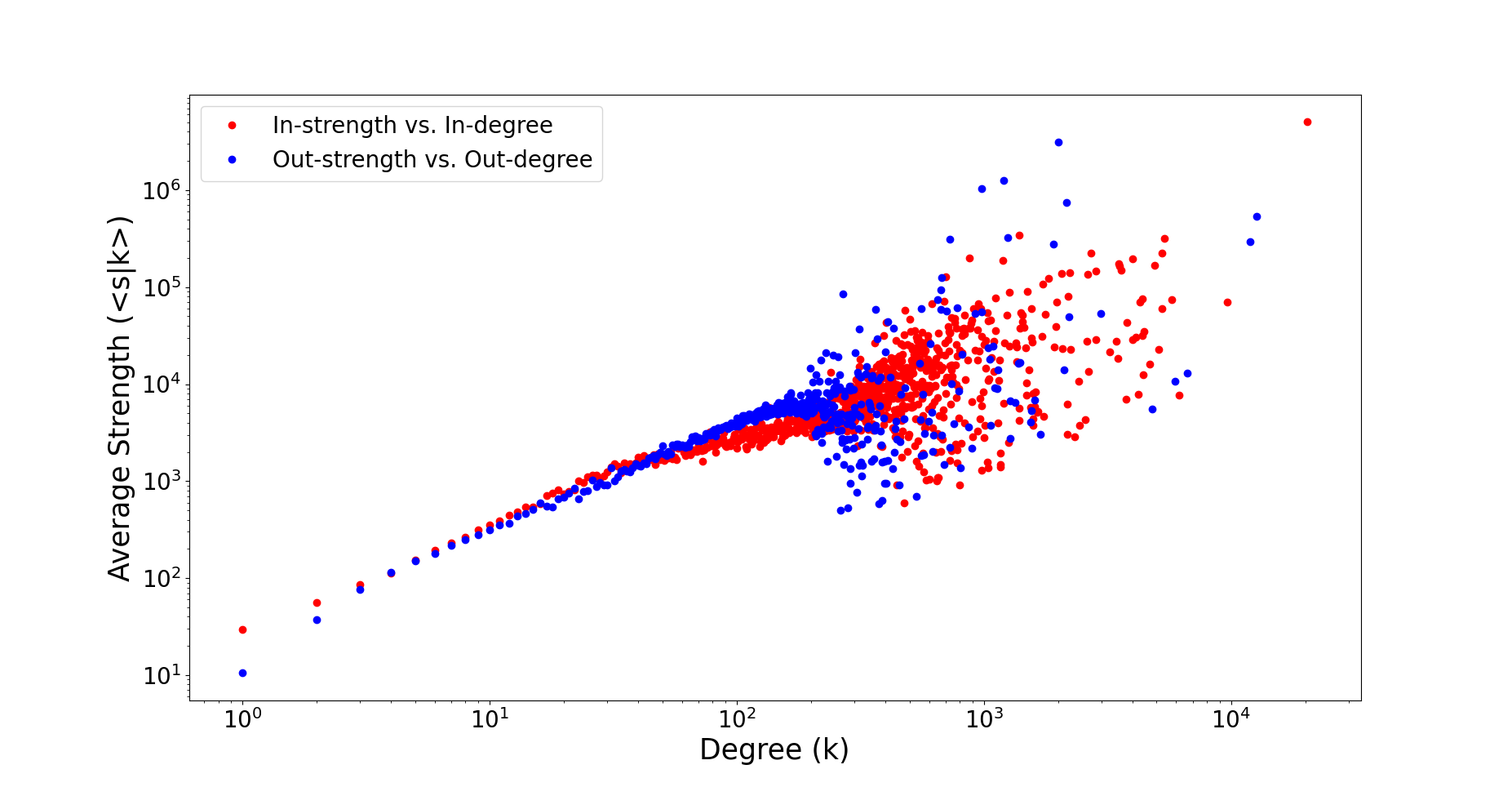}
    \caption*{(ii) $G^N$}
  \end{subfigure}
  \caption{Average In-strength vs. In-degree and average out-strength vs. out-degree.}
  \label{strvsdegree}
\end{figure}

We further analyze the average In-strength vs. In-degree and average out-strength vs. out-degree for understanding the correlation of average edge-weights from lower to higher degree nodes. The results are shown in Fig.~\ref{strvsdegree}. We observe that the average in/out-strength increases with in/out-degree till a certain range ($\sim 100$), and after that, no clear correlation is observed in $G^T$ network. The reason is that a user may have incoming or outgoing transactions with many users, but the transactions with those users might be performed for a small amount. For example, a user providing a paying-guest facility may have many incoming transactions, but each user will transfer a small amount as it is rare that a person stays at one place quite often. Due to different types of accounts having a higher degree, no clear correlation is observed for the higher degree nodes. However, the average in/out-strength increases with in/out-degree in $G^N$ network and has a higher correlation.

\subsection{Neighborhood Analysis}

For further understanding, we analyze the characteristics of the local neighborhood of nodes. We first compute the assortativity of degree and strength of the network that measures the similarity of connections in the network with respect to the node degree or strength, respectively \cite{newman2002assortative, newman2003mixing}. 
The assortativity values are mentioned in Table \ref{assortativitytable}, and neither degree or strength assortativity nor dis-assortativity is observed. These results are not in correlation with the assortativity observed in other scale-free networks, such as phone call based communication network \cite{onnela2007analysis} or the Internet network \cite{pastor2001dynamical}.

\begin{figure}[]
    \centering
    \includegraphics[width=\linewidth]{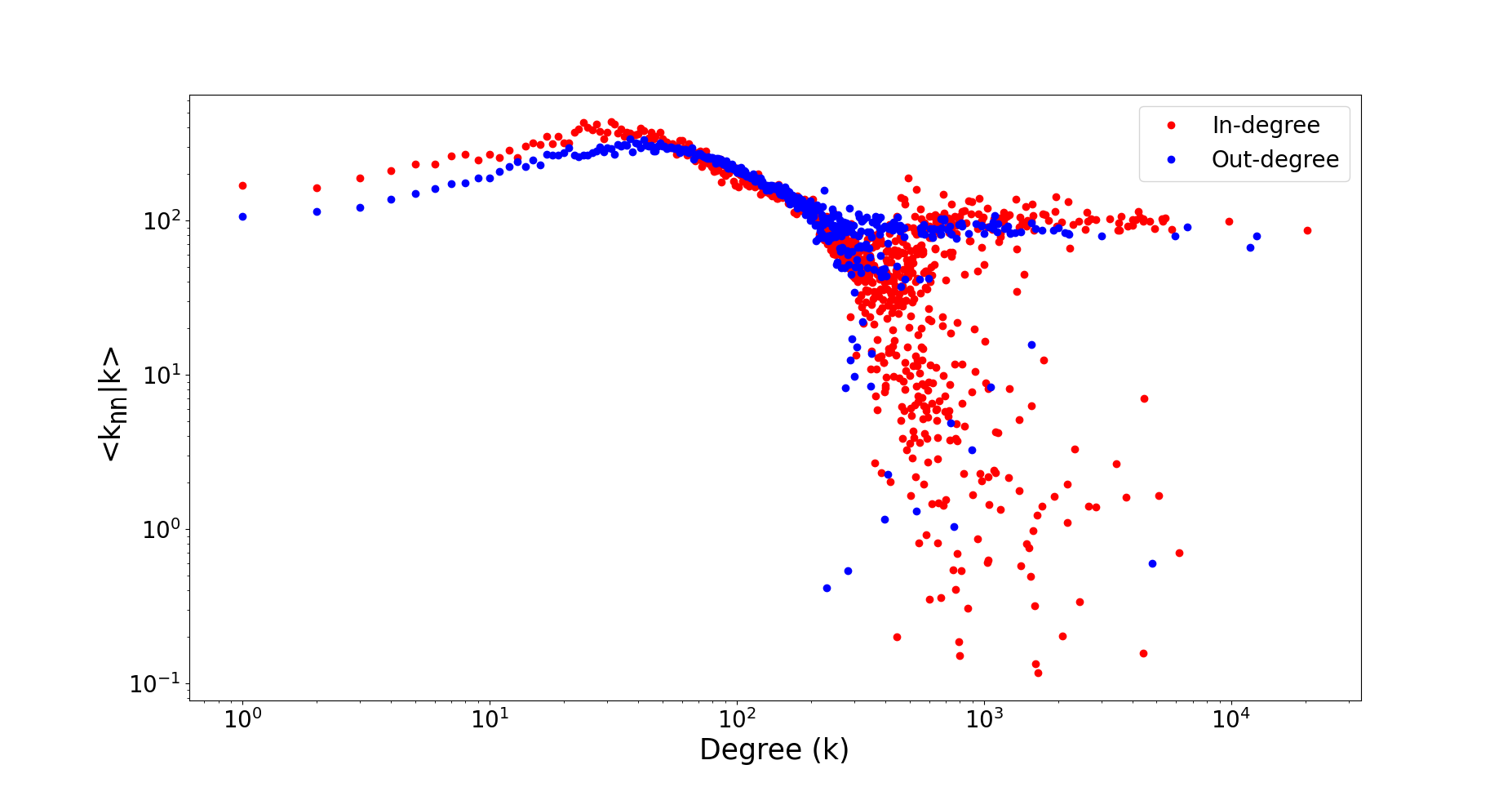}
    \caption{Average neighbor In-degree versus In-degree and average neighbor out-degree versus out-degree.}
    \label{avgnghdegree}
\end{figure}

\begin{figure}
  \begin{subfigure}{0.48\textwidth}
    \includegraphics[width=\linewidth]{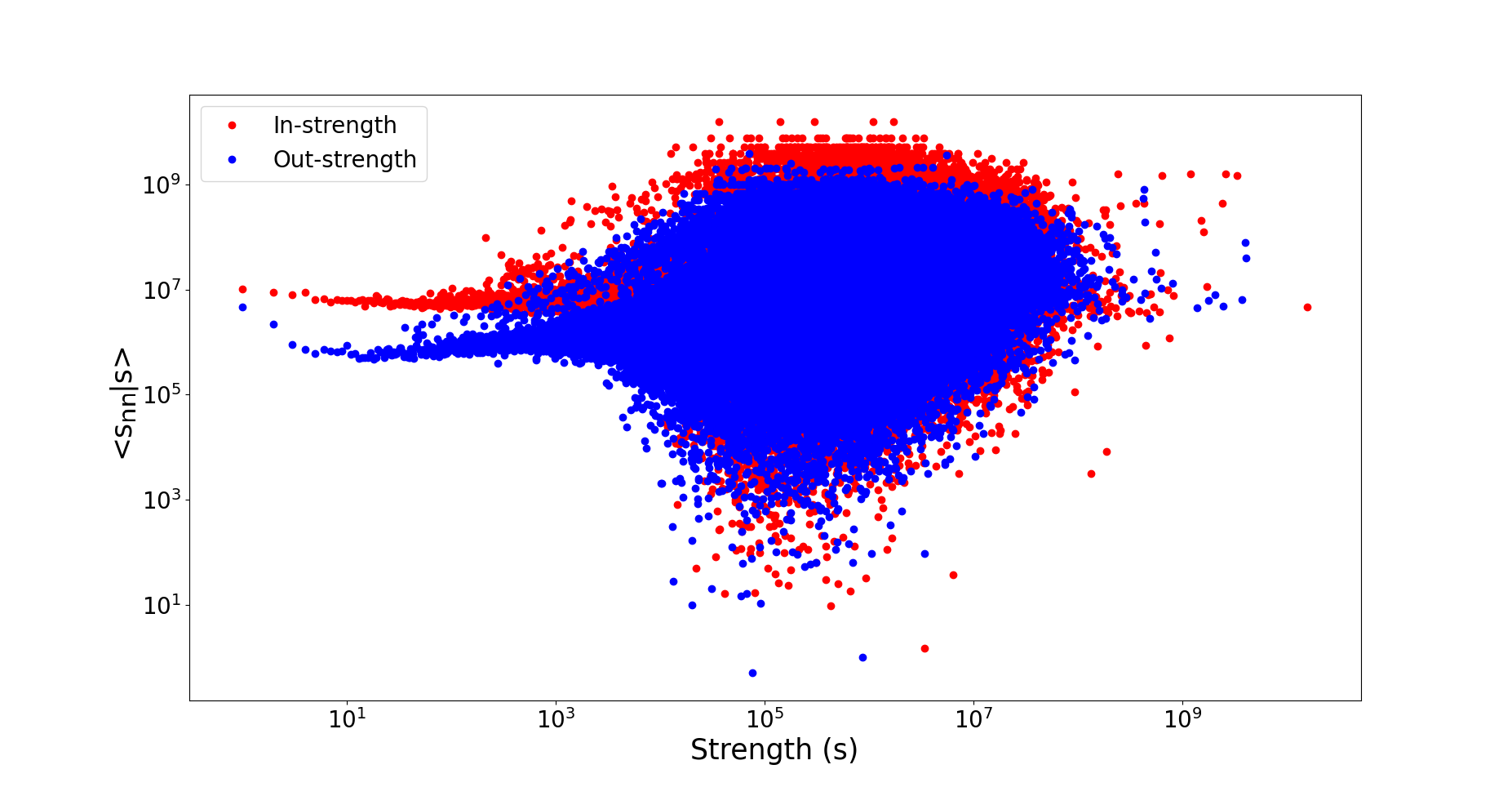}
    \caption*{(i) $G^T$}
  \end{subfigure}
  \hfill
  \begin{subfigure}{0.48\textwidth}
    \includegraphics[width=\linewidth]{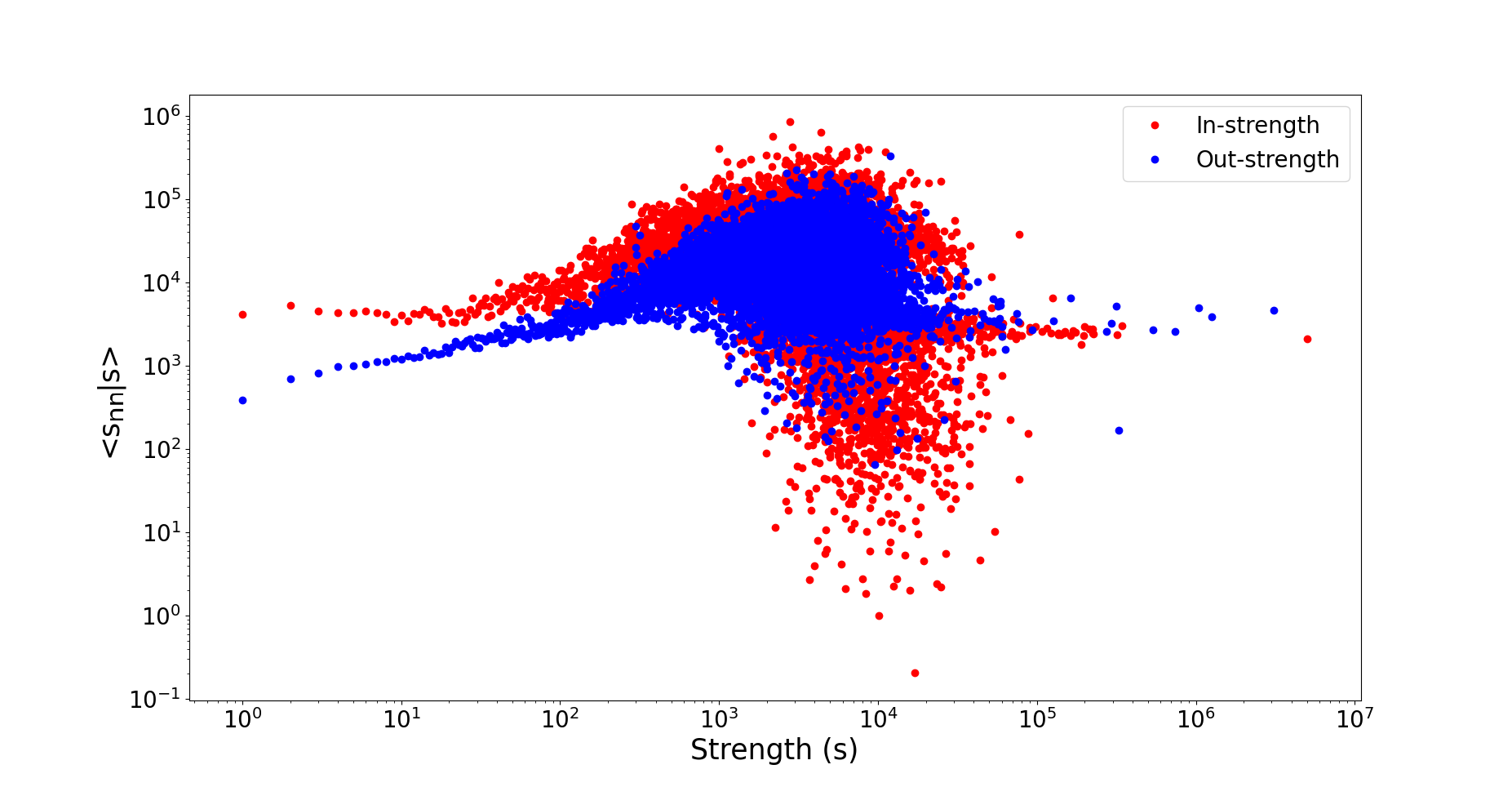}
    \caption*{(ii) $G^N$}
  \end{subfigure}
  \caption{Average Neighbor In-strength versus In-strength and average neighbor out-strength versus out-strength.}
  \label{avgneighborstrength}
\end{figure}

We plot the average neighbor In-degree versus In-degree $\left \langle k^{in}_{nn}|k^{in} \right \rangle$ and average neighbor out-degree versus out-degree $\left \langle k^{out}_{nn}|k^{out} \right \rangle$ in Fig.~\ref{avgnghdegree}, and observe that the most of the users have a high average neighbor degree irrespective of in/out-degree of a user. The average neighbor In-strength versus In-strength $\left \langle s^{in}_{nn}|s^{in} \right \rangle$ and average neighbor out-strength versus out-strength $\left \langle s^{out}_{nn}|s^{out} \right \rangle$ for $G^T$ and $G^N$ networks are shown in Fig.~\ref{avgneighborstrength} and no correlation is observed due to the basic network characteristics (as explained for Fig.~\ref{strvsdegree}). 

\begin{table}[]
\centering
\begin{tabular}{lr}
\hline
Assortativity                 & Value \\ \hline
In-degree Assortativity       & -0.018                    \\
Out-degree Assortativity      & -0.011                    \\
In-strength Assortativity $G^T$  & -0.004                    \\
Out-strength Assortativity $G^T$ & -0.005                    \\
In-strength Assortativity $G^N$  & -0.005                        \\
Out-strength Assortativity $G^N$ & -0.003    \\ \hline     
\end{tabular}
\caption{In and Out- Degree and Strength Assortativity values for different networks.}
\label{assortativitytable}
\end{table}

We compute the clustering coefficient of nodes that is the ratio of the number of closed triads in the neighborhood of a node with the total possible triangles. For unweighted networks, the clustering coefficient of a node $i$ is computed as \cite{fagiolo2007clustering},
\begin{equation}
    CC_i= \frac{1}{k^{tot}_i(k^{tot}_i-1) - 2k^r_i}T_u
\end{equation}
where $k^{tot}$ is the sum of in-degree and out-degree of node $i$, $k^r_i$ is the reciprocal degree of node $i$, and $T_u$ is the number of directed triangles through node $i$. 
For weighted networks, the clustering coefficient is computed as \cite{saramaki2007generalizations},
\begin{equation}
    CC_i= \frac{1}{k^{tot}_i(k^{tot}_i-1) - 2k^r_i}\sum_{t \in T^{dir}_{ijk}}(w_t)^{1/3}
\end{equation}
where $T^{dir}_{ijk}$ is the set of all directed triangle including node $i$, and $w_t$ for a triangle is computed as $\widehat{w_{ij}}\widehat{w_{jk}}\widehat{w_{ki}}$ if the triangle has edges $(i,j)$, $(j,k)$, and $(k,i)$ edges, where $\widehat{w_{ij}}$ is the normalized edge-weight by that is computed by dividing the edge-weight $w_{ij}$ by the maximum weight in the network.

\begin{figure}[t]
    \centering
    \includegraphics[width=\linewidth]{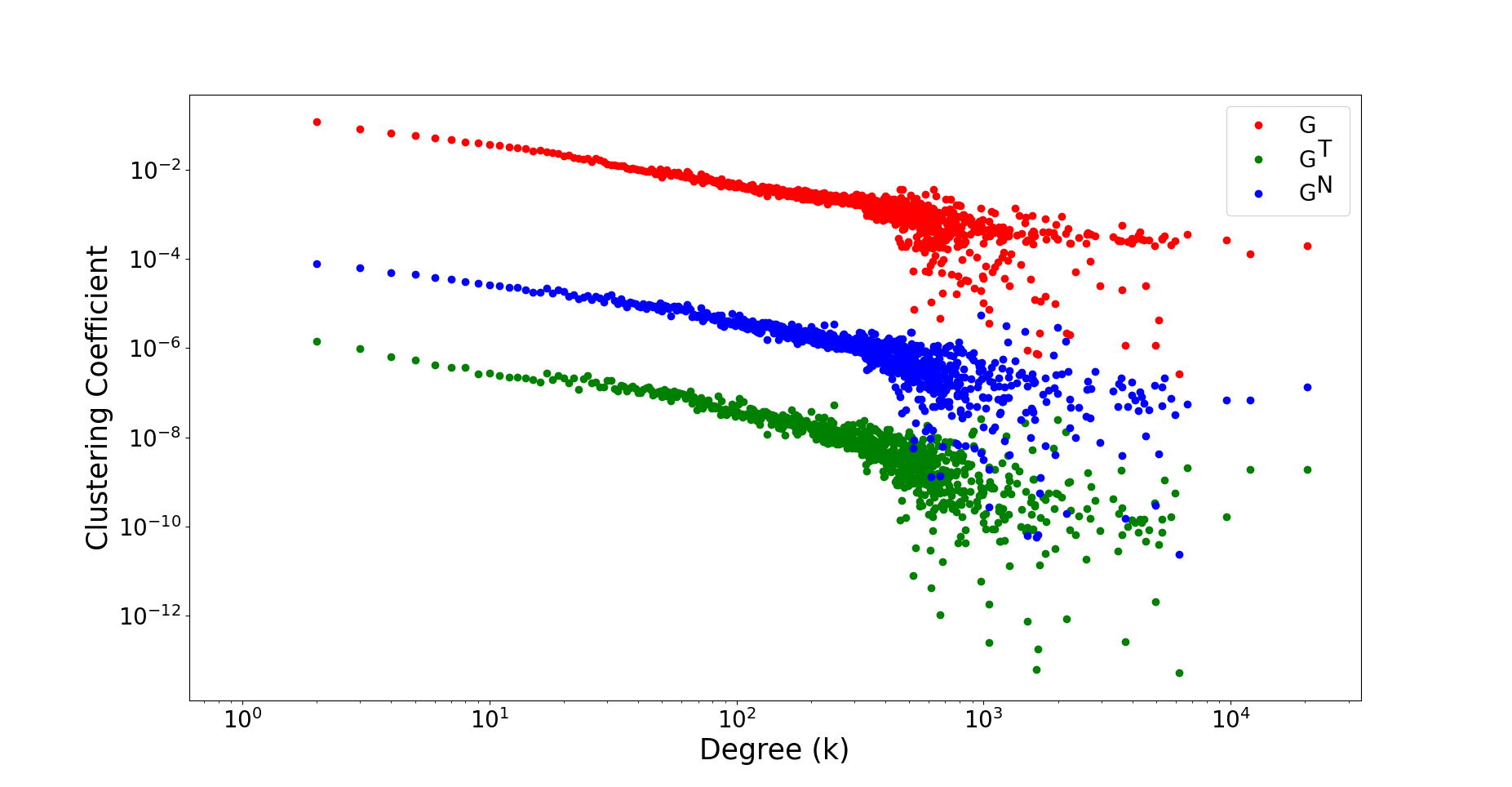}
    \caption{Average clustering coefficient versus total degree.}
    \label{avgclustering}
\end{figure}

The average clustering coefficient versus degree is plotted in Fig.~\ref{avgclustering}. The clustering coefficient decreases with the total degree of nodes. It shows that when a user makes more transactions with different accounts, it is less probable that the neighboring accounts will make a transaction with each other. The clustering coefficient follows the same pattern as observed in other large-scale scale-free unweighted and weighted networks \cite{onnela2007analysis, schiavo2010international}.

In Table \ref{networkstatistics}, we summarize the characteristics of the considered networks that include the minimum, maximum, mean, standard deviation, skewness\footnote{The skewness measures the asymmetry of the given probability distribution about its mean.}, and Kurtosis\footnote{Kurtosis computes how heavily the tail of the distribution differs from the tail of a normal distribution.} values of various basic network properties. 

\begin{table*}[]
\centering
\begin{tabular}{llrrrrrr}
\hline
                    &                                 & \multicolumn{1}{r}{Min} & \multicolumn{1}{r}{Max} & \multicolumn{1}{r}{Mean} & \multicolumn{1}{r}{Std} & \multicolumn{1}{r}{Skewness} & \multicolumn{1}{r}{Kurtosis} \\ \hline
\multirow{3}{*}{G}  & In-degree                       & 0                       & $2.04 \times 10^4$                   & 2.36              & 32.66            & $2.17 \times 10^2$             & $1.05 \times 10^5$                 \\
 & Out-degree                      & 0                       & $1.27 \times 10^4$                  & 2.36              & 20.26             & $3.35 \times 10^2$             & $1.84 \times 10^5$                 \\
 & Clustering Coefficient          & 0                       & 1                       & 0.02            & 0.12            & 7.29              & 53.95                  \\ \hline

\multirow{4}{*}{$G^T$} & In-strength                     & 0                       & $1.57 \times 10^{10}$             & $1.54 \times 10^5$              & $1.34 \times 10^7$            & $1.01 \times 10^3$               & $1.18 \times 10^6$                 \\
    & Out-strength                    & 0                       & $4.04 \times 10^9$             & $1.54 \times 10^5$             & $6.53 \times 10^6$            & $4.71 \times 10^2$             & $2.61 \times 10^5$                 \\
    & Edge-weight                     & 1                       & $3.35 \times 10^9$              & $6.56 \times 10^4$              & $2.75 \times 10^6$            & $8.99 \times 10^2$             & $9.43 \times 10^5$                  \\
 & Weighted Clustering Coefficient & 0                       & $2.72 \times 10^{-3}$          & $2.31 \times 10^{-7}$                 & $7.30 \times 10^{-6}$               & $1.62 \times 10^2$             & $4.22 \times 10^4$                 \\ \hline

\multirow{4}{*}{$G^N$} & In-strength                     & 0                       & $5.04 \times 10^6$             & $68.16$             & $4.08 \times 10^3$             & $1.17 \times 10^3$             & $1.44 \times 10^6$                  \\
 & Out-strength                    & 0                       & $3.11 \times 10^6$             & $68.16$             & $2.94 \times 10^3$             & $8.19 \times 10^2$              & $8.01 \times 10^5$                  \\
& Edge-weight                     & 1                       & $2.58 \times 10^4$              & $28.93$             & $1.19 \times 10^2$             & $34.02$              & $3.24 \times 10^3$                 \\
 & Weighted Clustering Coefficient & 0                       & $5.47 \times 10^{-2}$        & $1.41 \times 10^{-5}$                & $1.83 \times 10^{-4}$               & $71.15$              & $1.24 \times 10^4$             \\   \hline
\end{tabular}
\caption{Summary of descriptive network statistics.}
\label{networkstatistics}
\end{table*}

\section{Meso-scale Network Characteristics}\label{secmeso} 

In this section, we analyze the meso-scale characteristics of the network. The macro-scale characteristics focus on the properties of the network, and micro-scale characteristics analyze the properties of the nodes; however, meso-scale characteristics mainly focus on the structure of the network. For example, how the nodes are connected, the evolution was random or regulated by any evolving phenomenon, and so on.

First to analyze the local structure of the network, we compare the presence of cliques of different orders as these are considered the basic building blocks and moderate different functionalities of the network \cite{milo2002network, kashtan2004topological}. We compare the cliques of order $k = 1, 2,...,8$ with random networks in which the probability to have a connection is the same for all pairs of node. The random networks are created using Erd\H{o}s-R{\'e}nyi (ER) network generative model \cite{erdos1960evolution}. The link formation probability in the ER graph is computed as, $p = 2m/(n(n - 1))$, and the expected number $E[X]$ of cliques with $k$ nodes is computed as $E[X] = \begin{pmatrix}n\\  k \end{pmatrix} \left ( k!/a \right ) p^l$, where $l = k(k - 1)/2$ and $a = k!$ is the number of automorphic graphs, those are isomorphic to node-edge order preserved permutation of the graph \cite{bollobas2001random}. The cliques are counted by considering the undirected version of the network, and the results are shown in Table \ref{noofcliques}. Note that $k = 1$ corresponds to the number of nodes $n = 1622173$ and $k = 2$ to the number of edges $m = 3318903$, which are the same in the Rabobank and random network. The presence of high-order cliques (cliques beyond order three) in the Rabobank network makes it starkly different from a random network that has a very low probability of high-order cliques being present.

\begin{table}[]
\centering
\begin{tabular}{lll}
\hline
Order & Empirical Count & ER Expectation \\ \hline
1     & 1622173         & 1622173        \\
2     & 3318903         & 3318903        \\
3     & 417461          & 11.42              \\
4     & 34638           & $7.43 \times 10^{-11}$             \\
5     & 3815            & $9.76 \times 10^{-28}$              \\
6     & 417             & $2.70 \times 10^{-50}$              \\
7     & 26              & $1.61 \times 10^{-78}$             \\
8     & 0              & $1.12 \times 10^{-112}$ \\ \hline
\end{tabular}
\caption{Number of cliques of order $k = 1, 2,...,8$ in the undirected Rabobank network (empirical count) and their expectation values in a corresponding ER network (ER expectation).}
\label{noofcliques}
\end{table}


We also compare the reciprocity \cite{ruzzenenti2010complex} of the network with random networks. The reciprocity of a directed network is defined as the ratio of the number of edges pointing in both directions to the total number of edges in the network; it is computed as, $r=\frac{|(i,j) \in G|(j,i) \in G|}{|E|}$. 
The expected value of reciprocity in an Erd\H{o}s-R{\'e}nyi directed network is computed as $E[r]= \frac{m}{n(n-1)}$, where $n$ is the number of nodes and $m$ is the number of edges. The network $G$ has reciprocity $0.26$ that is much higher than the expected value of the reciprocity in a random network that is $1.05 \times 10^{-6}$.

To further understand how the users are connected in a group and how these groups are further connected with each other, we analyze the community structure of the network. For the Rabobank data, the information of ground truth communities is not known. However, the unavailability of the ground truth data is a well-known issue in analyzing network data, and the researchers have proposed several methods to identify communities using the structural properties of the network. We use the Leiden community detection method \cite{traag2019louvain} that is an extension of the Louvain community detection method to identify well-connected communities in directed weighted/unweighted networks. The Leiden algorithm starts from a singleton partition and executes in three phases (i) move nodes from one community to another to identify a partition, (ii) refinement of the partition, and (iii) create an aggregated network based on the refined partition using the non-refined partition. The algorithm is applied iteratively until there is no further improvement. The implementation for the community detection and their evaluation is used provided by \cite{rossetti2019cdlib}. 

\begin{figure}
\centering
    \includegraphics[width=\linewidth]{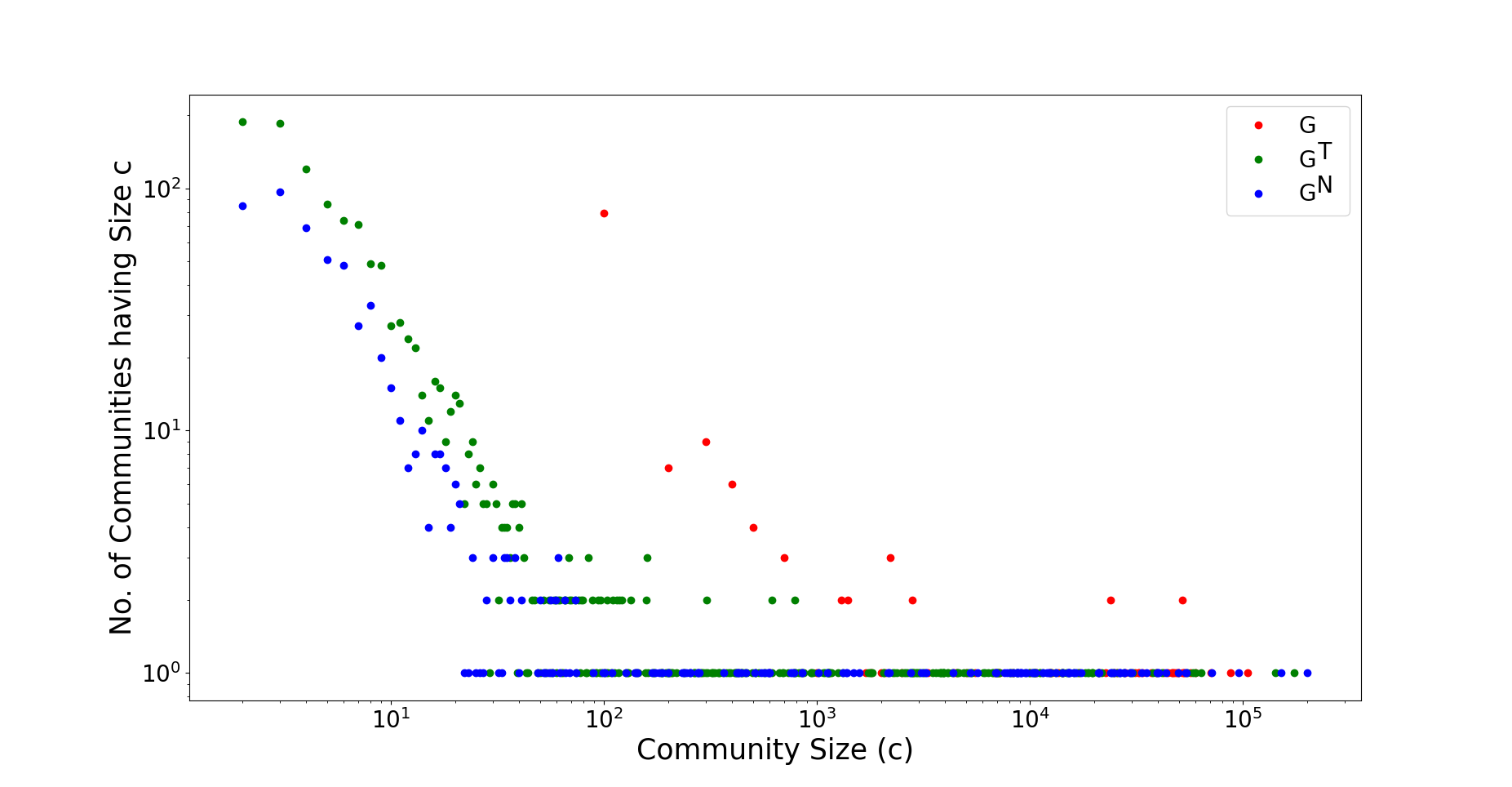}
  \caption{No. of Communities versus community size.} 
  \label{commdist}
\end{figure}

The distribution of the number of communities versus community size is shown in Fig.~\ref{commdist}. We observe that there are several small-size communities and a few big-size communities as observed in several scale-free networks \cite{arenas2004community}. The communities for network $G$ are linearly binned with bucket size 100. In Table \ref{communityevaluation}, we discuss the properties of the observed communities. The first row tells the number of communities; next, we discuss the min, max, mean and std. of community size. The communities are further evaluated using (i) Modular Density \cite{zhang2010determining} that explains the expected density of community, (ii) Hub Dominance \cite{lancichinetti2010characterizing} that defines the ratio of the degree of the highest connected node with respect to the theoretically maximal degree within the community, (iii) Conductance \cite{shi2000normalized} that computes the fraction of total edges that point outside the community, and (iv) Normalized-cut \cite{shi2000normalized} that is the normalized value of the fraction of existing edges (out of all possible edges) leaving the community. 
The results show that the bank transaction networks contain community structure as observed in other networks, such as social networks, co-authorship networks, financial networks, and so on \cite{arenas2004community}. However, the limitation of this analysis is that it is entirely based on the Leiden community detection method due to the unavailability of the ground-truth data.

\begin{table}[b]
\centering
\begin{tabular}{lrrr}
\hline
Metric/Network     & \multicolumn{1}{r}{G}   & \multicolumn{1}{r}{$G^T$}   & \multicolumn{1}{r}{$G^N$}   \\ \hline
\#Communities      & 219 & 1434 & 685 \\ \hline
Size-Min           & 3                       & 2                       & 2                        \\
Size-Max           & 129545	& 173698 &	200910                  \\
Size-Mean          & 7407.18 &	1131.22	& 2368.14                \\
Size-Std.         & 17927.49 &	7456.09 &	12261.84                \\ \hline
Modularity Density & 418.99	& 1557.24 &	998.55   \\ \hline
Hub Dominance &	0.86 &	1.22 &	1.26   \\ \hline
Conductance	&	0.10 &	0.22 &	0.17 \\ \hline
Normalized-cut	&	0.10 &	0.22 &	0.17   \\ \hline
\end{tabular}
\caption{Characteristics of communities identified using Leiden Community Detection method.}
\label{communityevaluation}
\end{table}

\begin{figure}[t]
    \centering
    \includegraphics[width=\linewidth]{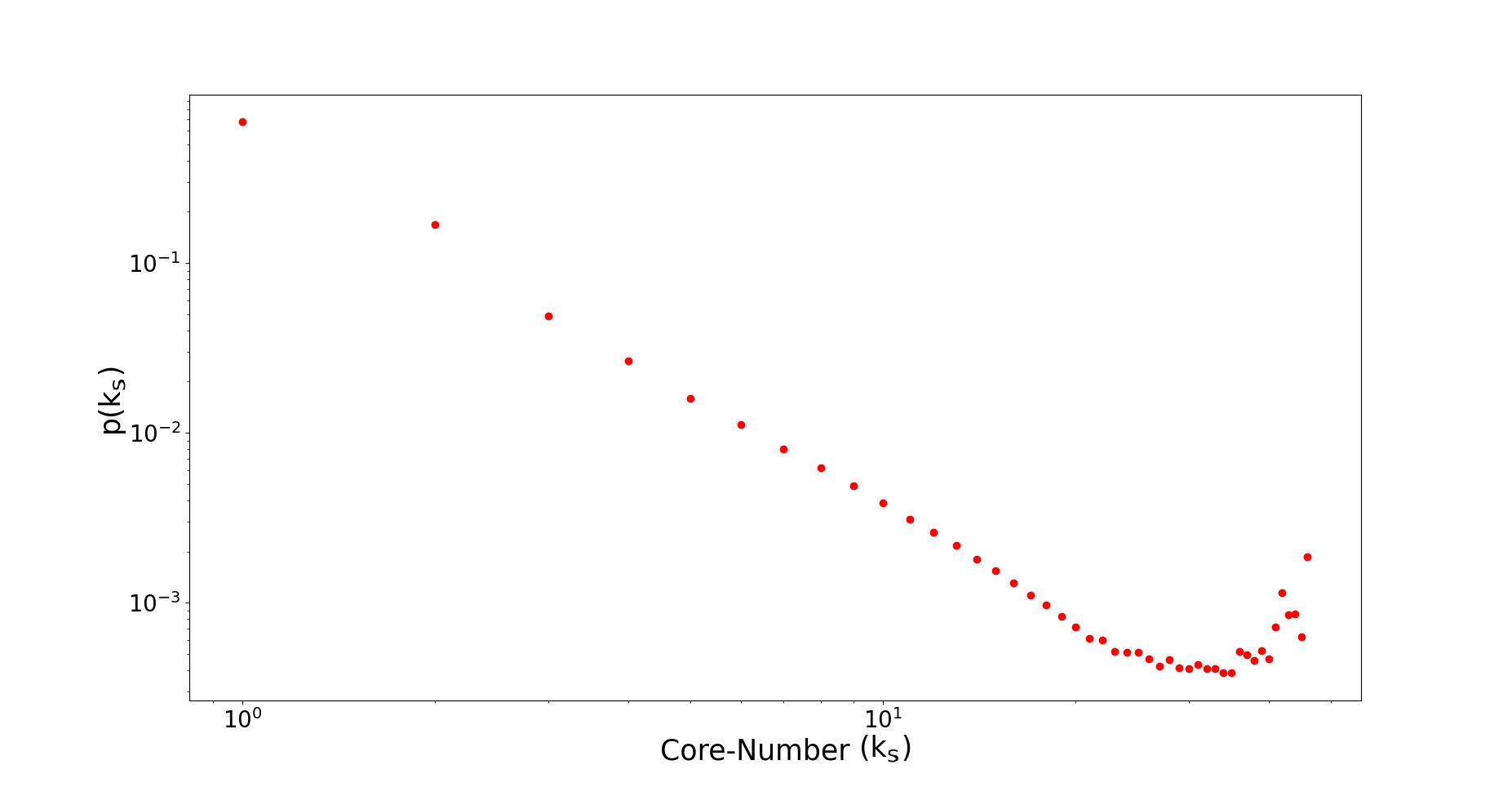}
    \caption{Distribution of nodes from periphery to core.}
    \label{coreperiphery}
\end{figure}

Apart from community structure, we analyze the core-periphery structure of the network that is another well-known meso-scale network structure \cite{borgatti2000models,saxena2016evolving}. In a banking transaction network, the core-periphery network will provide insights into the flow of money from peripheral to more central nodes. The core-Periphery structure is analyzed using k-shell decomposition algorithm \cite{seidman1983network}. In the k-shell method, first, we remove all nodes of degree (in-degree + out-degree) 1 until there is no node of degree one and assign them shell-index ($k_S$) 1. Then iteratively, we remove nodes of degree 2, 3, 4, ... until all nodes have a shell-index value. While removing nodes of degree $k$ if any node having degree less than $k$ appears, it is also removed in the same iteration and is assigned shell index $k_s=k$. In Fig.~\ref{coreperiphery}, we show the distribution of the nodes as we move from the outer peripheral layer to the inner core layer. The inner layers have a very small number of nodes. In network $G$, the highest shell-index is 46 shared by 2,871 nodes that is a very small fraction of all nodes. The results are similar to as observed in other networks where only a few nodes acquire the central position in a network \cite{saxena2016evolving}. 

The structural analysis of the network using k-clique, reciprocity, community, and core-periphery structure, shows that the network is not random and follows the structural properties observed in other real-world large-scale scale-free networks \cite{onnela2007analysis, tsai2005characteristics, saxena2020survey}.

\section{Characteristics of Nodes and Edges}\label{seccentrality} 

\begin{figure}[b]
    \centering
    \includegraphics[width=\linewidth]{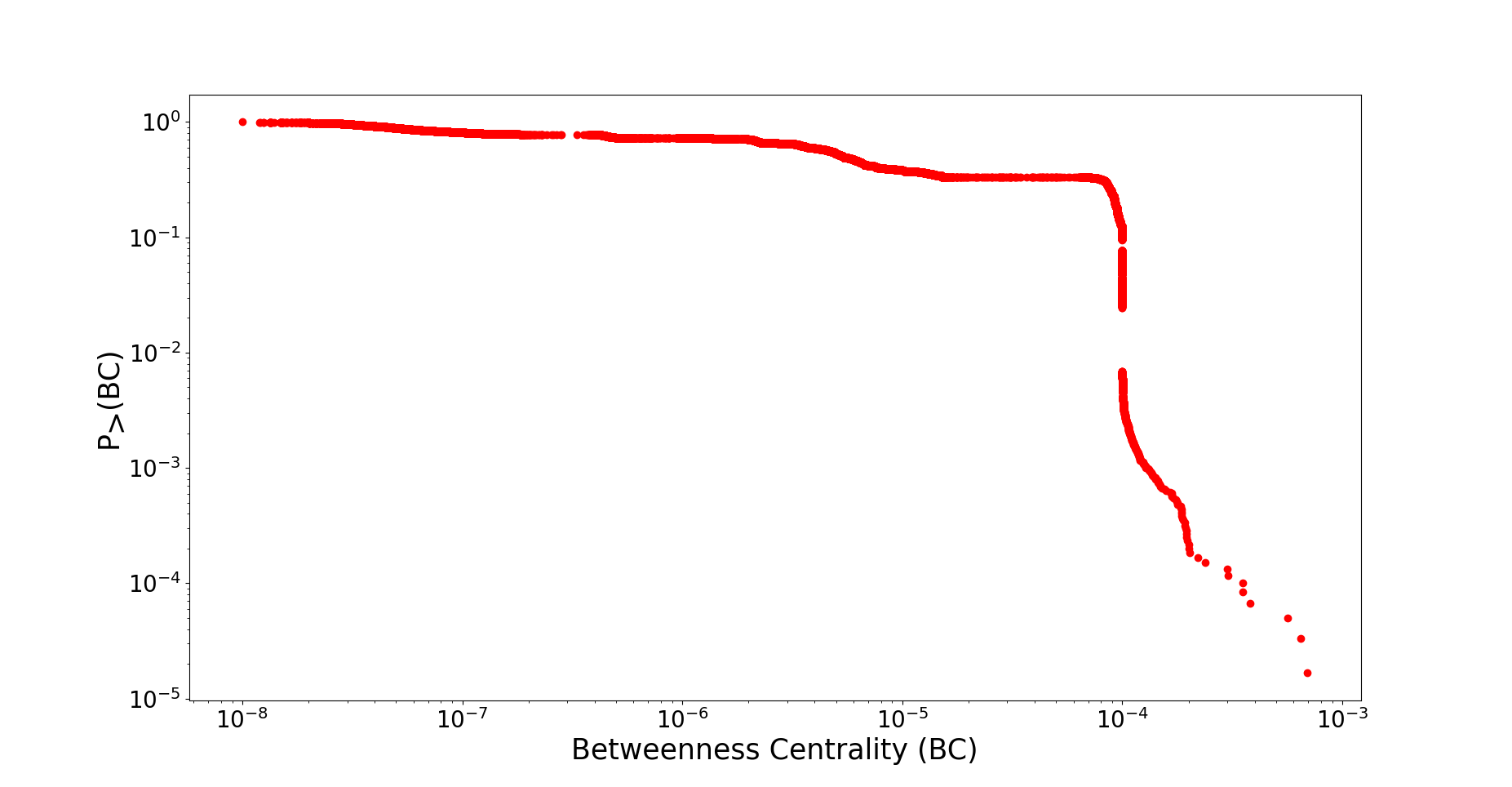}
    \caption{The cumulative edge betweenness centrality for the sampled graph having 10k nodes.}
    \label{bc10kgraph}
\end{figure}

We briefly analyze the characteristics of nodes and edges using centrality measures \cite{saxena2020centrality} to observe their behavior in correlation with the scale-free network structure. We first analyze the betweenness centrality of edges to study the role of a transaction in the money flow. The betweenness centrality of an edge is computed based on the number of shortest paths passing through that edge \cite{brandes2008variants}. The algorithm is computationally costly, we therefore u.a.r. sample a subgraph of 10,000 nodes having 59832 edges and show the cumulative distribution of the link betweenness centrality in Fig.~\ref{bc10kgraph}. The results show that it is very less probable to have a high betweenness centrality, confirming that most of the transactions happen within groups and a few transactions happen with the rest of the network.

\begin{figure}[t]
    \centering
    \includegraphics[width=\linewidth]{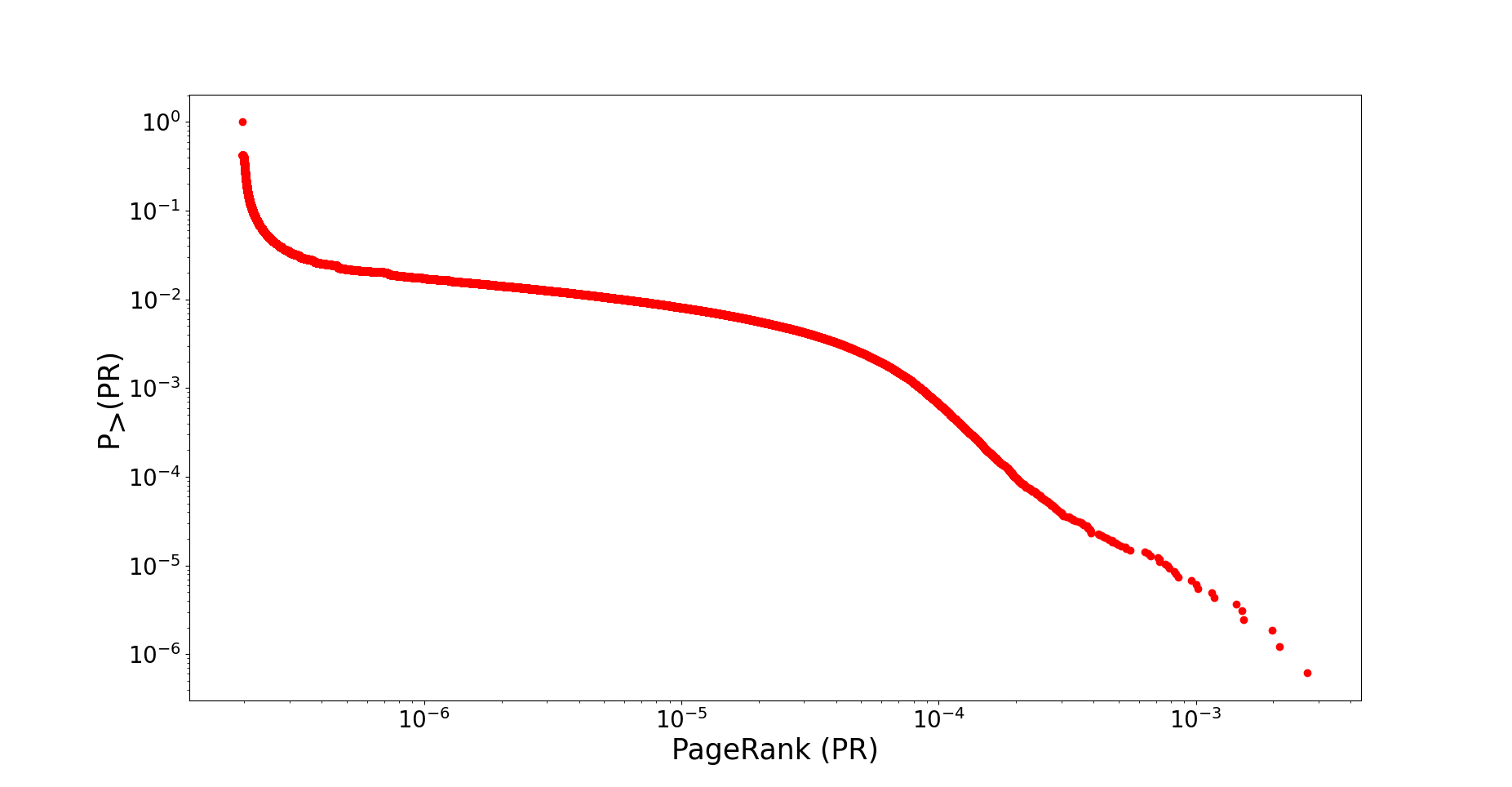}
    \caption{The cumulative PageRank distribution on network $G$.}
    \label{pagerank}
\end{figure}

To analyze the importance of nodes, we compute PageRank of network $G$, and its cumulative distribution is shown in Fig.~\ref{pagerank}. The network has a few nodes having very high PageRank and many nodes having lower PageRank value, as also observed in other scale-free networks \cite{shafiq2013identifying, newman2006structure}. The experimental analysis of link betweenness centrality and PageRank concludes that a few transactions and a few nodes have much higher importance than others in the bank transaction network. The one limitation of this analysis is that the comparative study of nodes' and edges' characteristics with their attributes couldn't be performed due to the limited information available in anonymized dataset. 

\section{Transaction network vs. Scale-Free Networks}\label{seccompare}

The banking transaction network follows the characteristics of scale-free networks; still, it has some clear similarities and differences with other scale-free real-world networks. The degree distribution, strength distribution, edge-weight follows power law as observed in other social, information, biological or technological networks \cite{newman2006structure}; however, there is no correlation between in-degree vs. out-degree, and in-strength vs. out-strength as observed in other Information networks, such as WWW \cite{hu2009empirical}. In social and Information (transportation) networks, the strength of nodes increases with degree \cite{onnela2007analysis, barrat2004modeling}; however, the banking transaction network has different pattern due to the nature of the network evolution (please refer to Section \ref{secbasic}). These basic characteristics show that neighborhood connectivity is different from other kinds of networks. The network is neither assortative nor disassortative, as we observe in most of the other scale-free networks including trade and finance networks \cite{newman2002assortative, schiavo2010international}; therefore, no correlation with the neighborhood nodes' degree is observed. 

The evolution of the network is not random and regulated by an underlying evolving mechanism. The network has community, and hierarchical structure \cite{arenas2004community, saxena2016evolving} as observed in other scale-free networks as shown in Section \ref{secmeso}. One another main difference that we observed is the correlation of weak ties with edge-weights. In social networks, the weak ties, or also referred to as bridges, are the connections between communities and therefore have a lower edge-weight as the social interaction between people belonging to two different communities is not very strong \cite{onnela2007analysis}. However, in the banking transaction network, no such correlation is observed. The transactions that happened between users belonging to different communities might have lower or higher strength based on the transferred amount and the total number of transactions. The evolving model for banking transaction networks is still an open question, and the above statistical observations mentioning the similarities and differences will help pin down the underlying evolving mechanism.


\section{Conclusion}\label{conclusion}

In this work, we constructed an unweighted network from bank transaction records of Rabobank and two weighted networks where edge-weights are assigned using the aggregated amount of transactions and the total number of transactions from 2010 to 2020. The network is useful in understanding the flow of money in society as it is derived from one-to-one money transactions. To our knowledge, it is the first intra-bank transaction network studied so far and will be the first bank transaction dataset that will be publicly available. One of our aims was to explore the relationship between the network topology and the money flow based on the associated weights. The analysis of k-cliques and reciprocity shows that the network evolution is not random. To further analyze, we study the network's community structure that shows that the users are organized in the smaller groups making more transactions between them and fewer transactions outside the group.

We believe that this bank transaction network and our analysis of the network's characteristics can help solve some practical tasks in the financial domain. For instance, in anomaly detection, the deviation of anomalous users' transaction behavior from structural characteristics of normal users might help in the early detection of suspicious and anomalous users. One can further analyze the mutual bank transaction network having connections if both end-users make transactions to each other as it will work as a good proxy to understand the underlying trust-based social network and its impact on the money flow.


\bibliography{mybib.bib}
\bibliographystyle{unsrt}

\end{document}